# Current-driven transformations of the intermediate state patterns in type-I superconductors

Jacob R. Hoberg and Ruslan Prozorov*

Ames Laboratory and Department of Physics & Astronomy,
Iowa State University, Ames, IA 50011

Dynamic structure of the intermediate state was studied in pinning-free thick Pb strips using real-time magneto-optical visualization. It is found that topological hysteresis can be lifted by applying sufficiently large current. Namely, laminar structure that appears on flux exit in a static case is turned into tubular when the sufficiently large transport current is present. Size and distribution of the flux tubes in static and dynamic regimes are different. Temperature, magnetic field and current phase diagram is discussed.

Over the years, much work has been done studying the intermediate state in type-I superconductors, lead (Pb) in particular [1,2,3]. Most of the research was performed on thin films, but the degree of pinning was rarely checked or attempted to be minimized (note that chemical purity does not imply the absence of pinning). While these studies provided some insight into the intermediate state structure, the question of the topology of the intermediate state remained open. It turns out that pinning is of utmost importance in defining the structure of the intermediate state [4,5]. While Landau laminar pattern has been observed in many experiments, on many occasions and in various materials direct observations revealed tubular structure [1,6]. Unlike Abrikosov vortices in type-II superconductors each bearing a single flux quanta, tubes in type-I superconductors may contain up to approximately $10^7$ flux quanta. Goren and Tinkham (GT) published a model in which flux tubes were considered as building blocks of the intermediate state [7]. It has never been confirmed experimentally citing various reasons. In our experiments with a thick (1.0 mm) Pb single crystal with very low pinning,

parameter–free, excellent agreement with the GT prediction was found. Figure 1 shows direct confirmation of the GT model from our measurements by analyzing the mean tube diameter as the function of a magnetic field. The solid line is a plot (not a fit) of the GT equation for the tube diameter, $D = [2\delta \cdot d /[(1-h)(1-h^{1/2})]]^{1/2}$, where $d$=1.0 mm is the sample thickness, $h=H/H_c$ is the reduced field and $\delta$ is the wall energy parameter [1]. For the plot, $\delta$ was set equal to the experimental value of 80 nm.

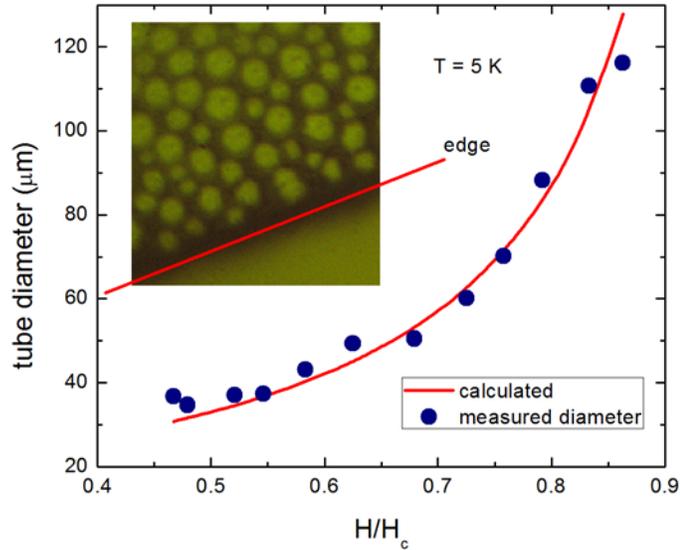

Figure 1. Mean tube diameter as a function of the applied field in Pb crystal measured at T = 5 K. The solid line is the Goren and Tinkham model [7] (not a fit) with parameters described in the text. Inset: example of flux tubes (bright) penetrating the sample from the edge marked by a line.

In recent papers, we showed that the intermediate state in thick samples with a rectangular cross-section exhibits a distinct behavior – tubes upon flux penetration and laminae upon flux exit [4,5]. We called this effect "topological hysteresis" to reflect the fact that hysteresis in the topology results in hysteresis in a macroscopically measurable magnetic moment. The tubular phase is robust and was shown to represent the equilibrium pattern by studying samples without geometric barrier [4] and it behaves as conventional 2D froth [8]. The laminar pattern forms due to continuity of the magnetic flux leaving the sample. The free energy difference is not sufficient to turn this pattern into tubes – it only leads to significant corrugation [9,10] of the initially straight lines. Perhaps this can happen in a dynamic state when domain walls are forced to move. In short, the laminar structure is topologically constrained, while the tubular structure is topologically mobile.

This leads to a natural question of the dynamical characteristics of flux tubes. Dynamics of small tubes in thin films was analyzed in terms of the geometric barrier [1,11,12]. Experiments with fast ramping magnetic field (equivalent to current) were interpreted in terms of dynamical re-organization of laminar domains [13], but were in fact snapshots of fast moving tubes driven by the Lorentz force as observed in 70s by Solomon and Harris [14] and in 90s by Dutoit and Rinderer [15] by imaging the flux structure driven by applied current. Figure 2 also shows such behavior.

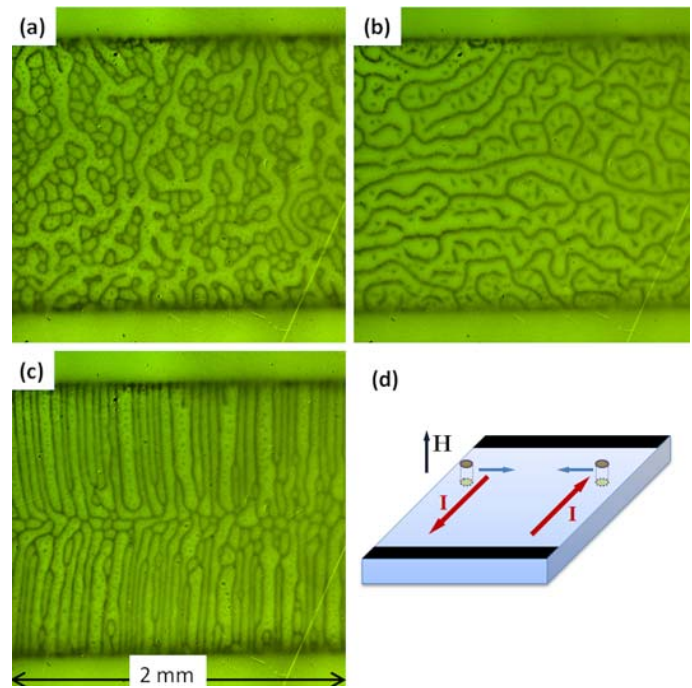

Figure 2. (a) flux penetration; (b) flux exit; (c) fast ramp penetration; (d) schematics of the experiment.

In this paper we discuss the influence of the transport current on the flux topology in the intermediate state of thick Pb strips. The prior works that involved transport current focused on the effects of geometric barrier [12] and estimation of flux velocity [14,16], whereas we are interested in topological transformations.

In the experiment, magneto-optical imaging of the component of the magnetic induction perpendicular to the surface was conducted by utilizing the Faraday effect in bismuth-doped iron garnet indicators with in-plane magnetization [17]. A flow-type liquid $^4$He cryostat with sample in vacuum was used. The sample was positioned on top of a copper cold finger and an indicator was placed on top of the sample. The cryostat was positioned under polarized-light reflection microscope and the color images could be recorded on video and high-resolution CCD cameras. When linearly polarized light passes through the indicator and reflects off the mirror sputtered on its bottom, it picks up a double Faraday rotation proportional to the magnetic field intensity at a given location on the sample surface. Observed through the (almost) crossed analyzer, we recover a 2D image [18]. Note that some images contain tooth-shaped overlay of darker and brighter areas. This is a side effect of a birefringence in the magneto-optical indicator that has in-plain magnetic domains and at certain angles with respect to the polarization plane these domains also show on the images. This, however, does not affect the underlying image in any way. High purity (Puratronic, 99.999%) Pb foils of various thicknesses were obtained from AlfaAesar. (In our experiments, Goodfellow foils showed much larger magnetic hysteresis and pinning as well as disordered patterns of the intermediate state).

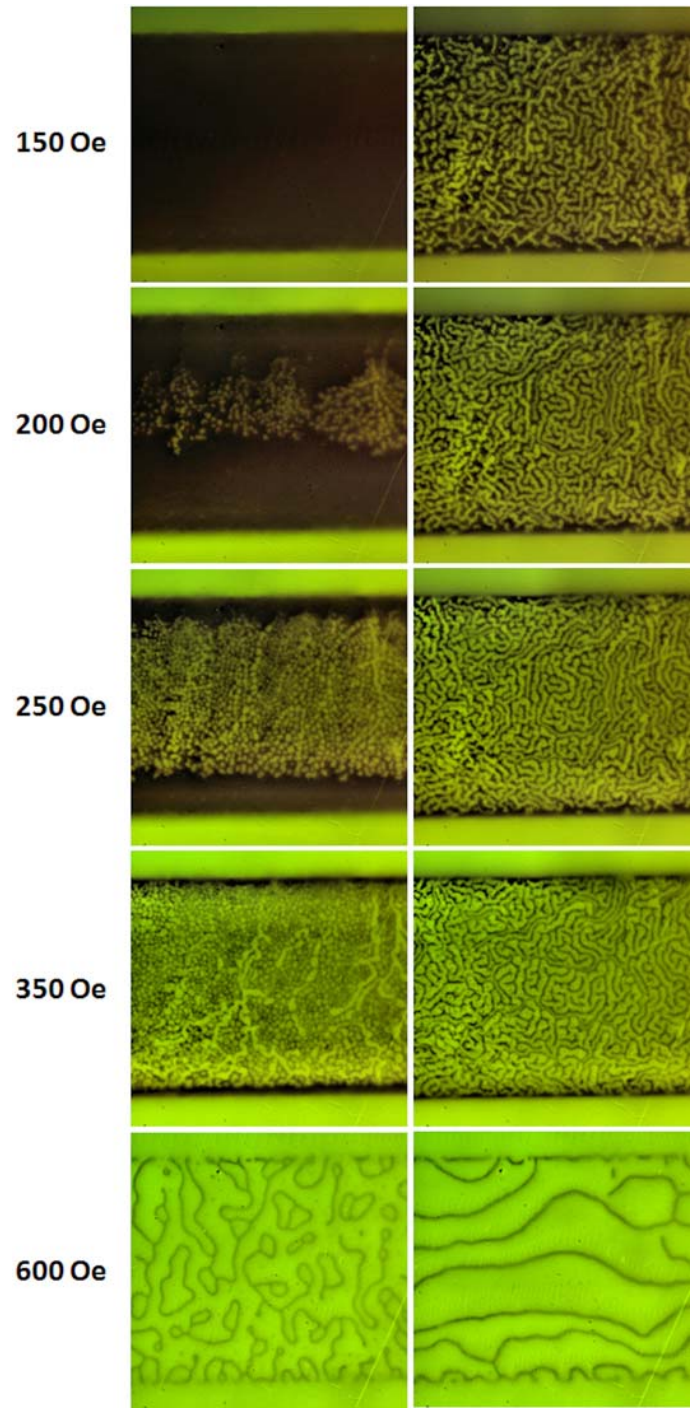

Figure 3. Pb strip of dimensions 9.30 mm x 1.85 mm x 0.48 mm zero field cooled with external field applied perpendicular to the area shown. The left column shows increasing fields (top to bottom). The right column shows decreasing fields (bottom to top).

Figure 3 shows magnetic flux penetration (left column) and exit (right column) at 4.8 K without applied current into a 5 mm long, 1.85 mm wide and 0.48 mm thick strip. As the applied magnetic field is increased, flux tubes "jet" into the center of the sample from the edges. These jets are caused by the Lorentz force exerted by the Meissner current on the tubes that just snapped off the finger-like protrusions of flux at the sample edge [19]. The tubes passing the geometric barrier region accumulate at the strip center. The tubular structure then, builds upon itself, slowly pushes outward towards the edge of the sample as the applied field is increased. Once the tubes reach the edge of the sample, they grow and eventually merge, increasing the normal phase area until the entire sample reaches the normal state. From this point, the applied field is decreased (right column, ascending), and the magnetic flux exits the strip in such a manner as to leave the Landau laminar structure.

With the applied field decreasing, the laminar structure stays uniform throughout the sample while merely "thinning" to allow flux exit and an increase in the area of the Meissner state. At small fields the laminar structure breaks at the sample edge and exits the sample as flux tubes [20].

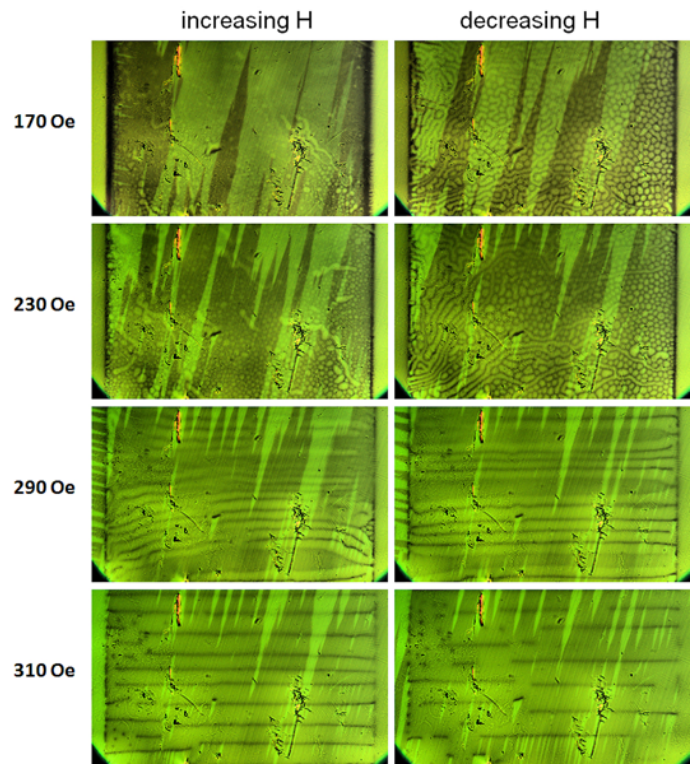

Figure 4. Pb strip of dimensions 9.10 mm x 2.15 mm x 0.25 mm zero field cooled with a constant current of 0.5 A applied. The left column shows increasing field (top to bottom). The right column

shows decreasing field (bottom to top). At fields 230, 290, and 310 Oe, there is constant motion of flux from the right side to the left. (See comment on the tooth-shaped overlay images).

An applied current in a Pb sample in the intermediate state will move the magnetic flux according to the Lorentz force. Figure 4 shows flux penetration and exit in a sample that was initially cooled in zero magnetic field to 5 K with a constant current of 0.5 A. Upon penetration, at small fields flux motion is slow and is perpendicular to the flow of current. At a certain applied field, the sample nucleates flux tubes at its edge. The nucleated flux tubes move slowly at first, but quickly gain speed. The highest velocity of which was measured at 3 mm/s. The magnetic flux has been found to move faster [1], but these higher speeds are not resolvable with our imaging equipment. At a high enough flux density, this dynamic structure switches over to a striped pattern. The stripe pattern can be seen to begin at 290 Oe and is fully developed at 310 Oe. Upon further increase of magnetic field the number of stripes decreases with individual stripes snapping off and creating what looks like a dislocation [20]. After bringing the sample to the normal state and subsequently lowering the applied field, this striped structure reappears. In stark contrast to a static case, the laminar structure does not develop. Instead, we see a tubular pattern very similar to that observed on flux entry (but with larger tubes).

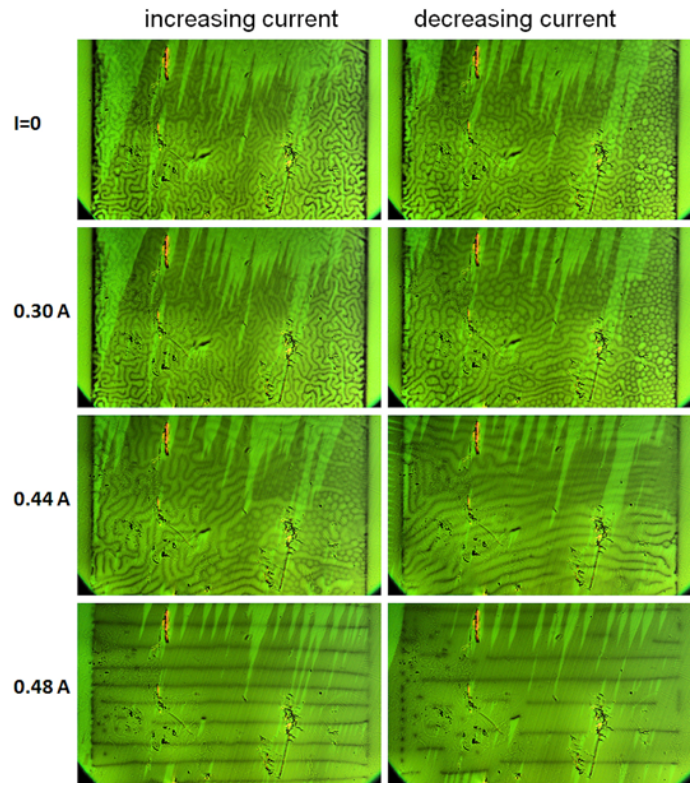

Figure 5. Pb strip of dimensions 9.10 mm x 2.15 mm x 0.25 mm field cooled in a 350 Oe applied magnetic field. The left column shows an increasing (top to bottom) applied current. The right column shows a decreasing (bottom to top) applied current after the sample was driven to the normal state using current.

In order to take a closer look at this response to current, two other basic experiments were performed. Figure 5 and Figure 6 show the situation when the applied field is fixed while the current that was applied is changed. Figure 5 shows a sample that has been field cooled in a constant applied field of 350 Oe. Figure 6 shows the experiment done after cooling in zero field after which the same 350 Oe field was applied. While the initial state in two cases is different, laminar in case of Figure 5 and tubular in case of Figure 6, the end result is very similar - tubular structure. It should be noted that tube size after ramping the current is much larger than that obtained after cooling in zero field.

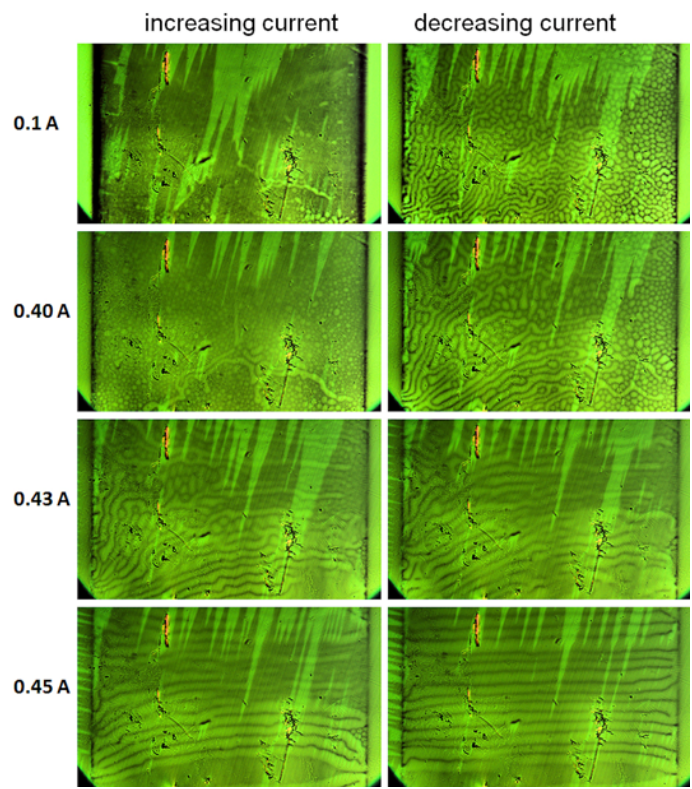

Figure 6. Similar to Figure 5, but after cooling in zero field.

Finally, by observing real-time imaging and recording characteristic events, a field vs. current diagram can be constructed. Figure 7 shows "tube nucleation" driven by field at constant current (filled squares). At an applied current of 0.35 A and below, the flux structure becomes too dense to create a full tubular motion and instead, at the higher fields moves directly into the "stripe pattern" phase. Above 0.35 A, the density of the flux tubes is low enough to allow motion of the internal field to the point of flux exit on one edge and the necessary flux tube nucleation at the other edge. At these higher currents, the stripe pattern phase is still available at higher fields. Also of note is the observation that the bubble nucleation line merges with the "1$^{st}$ motion" line at near-critical applied currents.

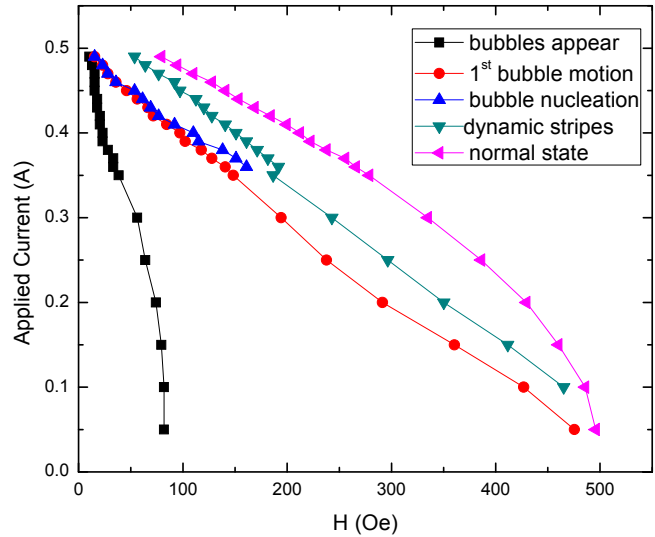

Figure 7. HI phase diagram showing: 1) first appearance of tubes 2) the start of motion 3) the start of bubble nucleation at the sample's edge 4) the formation of the striped flux structure 5) magnetic saturation.

In conclusion, presented experiments on a driven intermediate state demonstrate convincingly that tubular pattern represents a topologically equilibrium state. The laminar pattern is topologically constrained with many different configurations resulting in a multitude of the metastable states. The tubular structure on the other hand is uniquely defined and highly mobile. In a clean, pinning-free sample this gives advantage to the tubular structure, whereas pinning will stabilize the laminar and at higher values dendritic flux structure.

Acknowledgements: We thank Paul Canfield, John Clem and Vladimir Kogan for helpful discussions. This work was supported by the Department of Energy, Basic Energy Sciences under Contract No. DE-AC02-07CH11358, NSF Grant Number DMR-05-53285 and the Alfred P. Sloan Foundation.

*Corresponding author. prozorov@ameslab.gov